\documentstyle[12pt]{article}
\begin{document}
\normalsize
\centerline{\bf On calculation of effective galvanomagnetic 
characteristics of}
\centerline{\bf inhomogeneous metals. Exact solution for the 
longitudinal}
\centerline{\bf effective conductivity of polycrystals of} 
\centerline{\bf metals in high magnetic fields.}
\normalsize
\vskip 5mm
\centerline{\it Inna M. Kaganova}
\small
\centerline{Institute for High Pressure Physics Russian Academy of 
Sciences}
\centerline{142190 Troitsk, Moscow Region; e-mail: 
kaganova@hppi.troitsk.ru}
\vskip 3mm
\normalsize
\centerline{Moisey I. Kaganov}
\small
\vskip 3mm
\centerline{7 Agassiz Ave., Belmont, MA 02478, USA, e-mail:
MKaganov@compuserve.com} 
\normalsize

\begin{abstract}
In the framework of the perturbation theory an expression suitable for 
calculation of the effective conductivity of 3-D inhomogeneous metals in 
uniform magnetic field $H$ is derived. For polycrystals of metals with 
closed Fermi surfaces in high magnetic fields the perturbation series 
defining the longitudinal and the hall elements of the perturbation 
series can be summed allowing us to obtain the exact expression for the 
leading terms of all these elements of the effective conductivity 
tensor. 
\end{abstract}

\section{Introduction}

Calculation of the effective (macroscopic) conductivity of inhomogeneous 
media (in particular, of a randomly inhomogeneous medium) is one of the 
well-known problems drawing attention of scientists during a very long 
time. However, in the general case this problem has not been solved yet. 
In this Introduction we do not pretend to give the full list of 
references, but refer the readers to the review paper by A.G.Fokin 
\cite{1}, where the long list of references relating to different 
aspects of the problem is presented. We concentrate mainly on the known 
exact solutions for the effective conductivity tensor (ECT).

Exact solutions for effective characteristics of stochastically 
inhomogenous media can be found very rarely. With regard to calculation 
of ECT the simplest exact solution was obtained by Wiener as far as 1912 
(see Ref. \cite{1}). He examined an unbounded plane-parallel layered 
medium, where the conductive properties of a layer are described by the 
conductivity tensor for one of $N$ homogeneous isotropic components.

The other example is the Dykhne formula for ECT of some of two-
dimensional inhomogeneous media \cite{2}. Also for a plasma located in a 
strong magnetic field, in \cite{3} A.M.Dykhne obtained exact expressions 
for the effective conductivity and the effective Hall parameter, 
supposing that the density fluctuations are two-dimensional ones. The 
existence of these results is due to specific transformations allowed by 
the equations of 2-D problems. Recently S.A.Bulgadaev \cite{4} 
generalized the Dykhne formula obtaining new exact solutions for 2-D 
heterophase medium with an arbitrary number of phases $N$.

Some time ago A.M.Dykhne and I.M.Kaganova \cite{5,6} derived the exact 
solution for the effective surface impedance of an inhomogeneous metal 
in the frequency region of the local impedance (the Leontovich) boundary 
condition applicability. In this case the existence of the exact 
solution is caused by the presence of the physically meaningful small 
parameter, namely, the surface impedance of good metals. The result is 
valid both under conditions of normal and anomalous skin effect, 
allowing us to calculate effective impedance of strongly anisotropic 
polycrystalline metals \cite{6,7}. The inhomogeneity of the metallic 
surface can be due to the properties of the metal or/and surface 
roughness. The effective impedance of strongly curved one-dimensional 
metallic surface was discussed in \cite{8}.

Sometimes effective characteristics can be estimated qualitatively. For 
example, there are no regular methods allowing us to calculate the 
effective conductivity of 3-D polycrystals accurately, but some general 
reasoning suggested by Yu.A.Dreizin and A.M.Dykhne \cite{9} allowed them 
to estimate ECT of strongly anisotropic polycrystals.

When the exact solution for an effective characteristic cannot be found, 
perturbation theory calculations may be useful. The most accurate and 
physically meaningful method to take account of spatial fluctuations of 
the characteristic was developed by I.M.Lifshits and his co-workers. 
(The first paper related to this topic was \cite{10}). They proposed to 
start from equations of the problem coefficients of which are random 
functions of position. Averaging these equations one derives equations 
for the averaged fields, which allow him to determine the effective 
characteristic. This method is relatively simple, when the spatial 
fluctuations are small. Mostly it is used when only the first 
nonvanishing term of the perturbation series is taken into account. 
Usually, this term does not depend on the correlations of the stochastic 
characteristic in different points of the medium. 

In \cite{11} in the absence of a uniform magnetic field the perturbation 
theory formula, suitable when calculating the effective conductivity 
(EC) of 3-D inhomogeneous metals was derived. When the perturbation 
series converges, the result of \cite{11} allows one to calculate EC as 
the series in powers of invariants of the local conductivity tensor 
(LCT) ${\sigma}_{ik}({\bf r})$. EC of 3-D polycrystals was examined. The 
first term of the series taking account of spatial fluctuations of LCT 
is of the second order. As usual, if the polycystal is isotropic in 
average, this term does not depend on statistical properties of the 
inhomogeneous medium. However, beginning from the third order term the 
answer depends on the correlations of the fluctuating part of LCT in 
different points of the medium. 

In this paper we present the perturbation theory formula for the 
effective conductivity tensor (ECT) in a uniform magnetic field $\bf H$. 
We examine stochastically inhomogeneous metals that are isotropic and 
homogeneous in average. The way of deriving the expression for ECT is 
much similar to the algorithm proposed in \cite{11}. We apply the 
obtained result when calculating ECT of polycrystalline metals, composed 
of single crystal grains, supposing that the Fermi surface of the single 
crystal is a closed one. The case of very high magnetic fields ($r_c \ll 
l$, where $r_c$ is the cyclotron radius and $l$ is the electron mean 
free path) is examined. We show that under the abovementioned conditions 
the perturbation series relating to the longitudinal element and the 
Hall element of ECT can be summed, allowing us to calculate the leading 
terms of these elements of ECT exactly. The existence of this exact 
solution is due to the specific form of LCT when $r_c/l \ll 1$.

Classical galvanomagnetic properties of single crystal metals were 
scrutinized by a lot of authors. The results can be found in textbooks 
of electron theory of metals (see, for example, \cite{12}). In the 
general case the dependence of the elements of the single crystal 
conductivity tensor on the value and direction of the magnetic field 
cannot be obtained for an arbitrary value of $H$. The elements of the 
conductivity tensor depend on the details of the scattering 
processes, as well as on the dynamics of conduction electrons. However, 
the asymptotic behavior of the elements of the conductivity tensor in 
high magnetic fields is defined mainly by the structure of the Fermi 
surface of the metal.

The Fermi surfaces of real metals are extremely complex and differ 
significantly for different metals\cite{13}. The first paper where the 
magnetoresistivity of polycrystals of metals was calculated with regard 
to the presence of open electron orbits, was the work by I.M.Lifshitz 
and V.G.Peschanskii \cite{14}. In \cite{15} Yu.A.Dreizin and A.M.Dykhne 
discussed ECT of polycrystals of metals with open Fermi surfaces (of the 
"space mesh" and "corrugated cylinder" types). In this case in some 
specifically oriented grains conduction electrons move along open 
orbits. Due to the contribution of these grains to the transverse 
effective conductivity, its leading term being nontrivially dependent on 
$H$, is anomalously large. 

However, up to now ECT for polycrystals of metals with closed Fermi 
surfaces has not been examined. Our perturbation theory allows us to 
examine this type of polycrystalline metals. Note, even the shape of a 
closed Fermi surface can be extremely complex. We restrict ourselves 
with the case when the number of conduction electrons $n_e$ is not equal 
to the number of holes $n_h$. The case $n_e = n_h$, as well as weak 
uniform magnetic fields will be discussed elsewhere. 

The organization of this paper is as follows. In Section 2 with the aid 
of perturbation theory we derive an expression allowing us to calculate 
ECT of 3-D inhomogeneous metals in a uniform magnetic field. In Section 
3 we use the obtained result to examine ECT of polycrystals of metals 
with closed Fermi surfaces in high uniform magnetic fields. The obtained 
results are exact, when the single crystal anisotropy is not very strong 
and perturbation series converges.
As an example, we calculate the effective longitudinal conductivity for 
polycrystals of metals with an ellipsoidal uniaxial Femi surface. 
Concluding remarks are given in Section 4.

\section{Perturbation Theory for ECT of 3-D Inhomogeneous Metals in 
Uniform 
Magnetic Field}

If ${\bf H} \ne 0$, the elements of LCT ${\sigma}_{ik}({\bf H},{\bf r})$ 
are 
random functions of position vector $\bf r$, and in the same time they 
depend on 
the value and orientation of the fixed vector $\bf H$. By $<...>$ denote 
the 
ensemble average over all possible realizations of the medium. When 
${\bf H}=0$, 
the averaged conductivity $<{\sigma}_{ik}> = <\sigma>{\delta}_{ik}$, 
however, if 
${\bf H}\ne 0$, the averaged conductivity $<{\sigma}_{ik}({\bf H})>$ is 
a 
tensor, which does not reduce to the isotropic tensor ${\delta}_{ik}$ 
only. Let 
us write LCT as
$$
{\sigma}_{ik}({\bf H},{\bf r}) = <{\sigma}_{ik}({\bf H})> + 
{\Delta}_{ik}({\bf H},{\bf r}); \quad <{\Delta}_{ik}({\bf H},{\bf r})> = 
0,
\eqno (1)
$$
where ${\Delta}_{ik}({\bf H},{\bf r})$ are random functions of ${\bf 
r}$. The 
tensor ${\Delta}_{ik}({\bf H},{\bf r})$ is responsible for the spatial 
fluctuations of LCT.

By definition, ECT ${\sigma}_{ik}^{ef}$ is specified by equation
$$
<j_i> = {\sigma}_{ik}^{ef}<E_k>, \eqno (2)
$$
where $<{\bf j}>$ is the macroscopic direct current density and $<{\bf 
E}>$ is 
the uniform macroscopic electric field. Let ${\bf j}({\bf r})$ and 
${\bf E}({\bf r})$ be the local current density and the local electric 
field, 
respectively. They are solutions of the electrostatics equations 
$$
{\rm div}{\bf j} = 0, \quad {\rm rot}{\bf E} = 0, \eqno (3)
$$
with the material equation that is the Ohm law:
$$
{j}_{i}({\bf r}) = (<{\sigma}_{ik}({\bf H})> + 
{\Delta}_{ik}({\bf H},{\bf r})){E}_{k}({\bf r}). \eqno (4)
$$
Because of the tensor ${\Delta}_{ik}({\bf H},{\bf r})$, Eq.(3) 
constitute a system of stochastic equations for the local field 
${E}_{i}({\bf r})$.

We set ${\bf j}({\bf r}) = <{\bf j}> + \delta {\bf j}$ and 
${\bf E}({\bf r}) = <{\bf E}> + \delta {\bf E}$, where $<\delta {\bf j}> 
= <\delta {\bf E}>=0$. Substituting these expressions in Eq.(4) and 
averaging, we obtain 
$$
<j_i> = <{\sigma}_{ik}({\bf H})><E_k> + J_i; \quad J_i = 
<{\Delta}_{ik}({\bf H},{\bf r}){\delta E}_k>. \eqno (5)
$$
The uniform vector ${\bf J}({\bf H})$ describes the contribution of the 
spatial fluctuations of LCT to the elements of ECT.
 
Subtracting Eq.(5) from Eq.(4) we have
$$
{\delta j}_i = <{\sigma}_{ik}({\bf H})>{\delta E}_{i} + 
{\Delta}_{ik}({\bf H},{\bf r})<E_k> + 
D_i({\bf H},{\bf r}), \quad
D_i({\bf H},{\bf r}) = {\Delta}_{ik}({\bf H},{\bf r})
{\delta E}_{k} - J_i({\bf H}). \eqno (6)
$$
Evidently, $<\bf D> = 0$. The components of the vector $\bf D$ are at 
least quadratic in powers of the elements of the tensor ${\Delta}_{ik}$. 
When calculating ECT up to the first nonvanishing term taking account of 
spatial fluctuations, the vector $\bf D$ has to be omitted. However, 
just this vector defines corrections of higher orders.

From the electrostatic equations (3) it follows that the Fourier 
coefficient of the stochastic electric field ${\delta E}_{i}({\bf r})$ 
is
$$
{\delta E}_{i}({\bf k}) = -{q}_{ik}({\Delta}_{jk}({\bf H},{\bf k})<E_k>+ 
D_j({\bf H},{\bf k})), \eqno (7.a)
$$
where ${\Delta}_{jk}({\bf H},{\bf k})$ and $D_j({\bf H},{\bf k})$ 
are the Fourier coefficients of the elements of the stochastic 
tensor ${\Delta}_{jk}({\bf H},{\bf r})$ and the components of the 
stochastic vector $D_i({\bf H},{\bf r})$, respectively, and the 
tensor ${q}_{ik}$ is
$$
{q}_{ik} = \frac{k_ik_k}{k^2Q}, \quad 
Q = <{\sigma}_{ik}({\bf H})>\frac{k_ik_k}{k^2}, \eqno (7.b)
$$ 

We seek ${\delta E}_{i}$ and $J_i$ as series in powers of the elements 
of the tensor ${\Delta}_{ik}$: 
$$
{\delta E}_{i}({\bf r}) = \sum_{n=1}^{\infty}
{\delta E}_{i}^{(n)}({\bf r}); \quad J_i = 
\sum_{n=2}^{\infty}J_i^{(n)}. \eqno(8) 
$$
With respect to the definition of ${\bf J}$, we have $J_i^{(n)} = 
<{\Delta}_{jk}({\bf H},{\bf r}){\delta E}_{i}^{(n-1)}>$. Next, from 
Eq.(7.a) it follows that the Fourier coefficients of ${\delta 
E}_{i}^{(n)}({\bf r})$ are
$$
{\delta E}_{i}^{(1)}({\bf k}) = -{q}_{ij}{\Delta}_{jk}({\bf 
H},{\bf k})<E_k> \quad {\rm and}\quad
{\delta E}_{i}^{(n)}({\bf k}){|}_{n>1} = -{q}_{ij}D_j^{(n)}({\bf 
H},{\bf k}), \eqno (9)
$$
where $D_j^{(n)}({\bf H},{\bf k})$ is the Fourier coefficient of 
$D_j^{(n)}({\bf r}) = {\Delta}_{jk}({\bf H},{\bf r})
{\delta E}_{k}^{(n-1)}({\bf r}) - J_j^{(n)}({\bf H})$. 

We use Eq.(9) to write ${\delta E}_{i}^{(n)}({\bf r})$ as a sum 
of the lower-order terms. Indeed, suppose $n \ge 2$. Then with 
regard to the definition of the vector $D_j^{(n)}({\bf r})$ 
we have
$$
{\delta E}_{i}^{(n)}({\bf r}) = -\left\{\int\int {\rm d}^{3}k_{n}
{\rm d}^{3}k_{n-1}{q}_{ij_n}^{(n)}({\bf H}){\Delta}_{j_nk_{n-1}}
({\bf H};{\bf k}_{n}-{\bf k}_{n-1}){\delta E}_{k_{n-1}}^{(n-
1)}({\bf k}_{n-1})
{\rm e}^{i{\bf k}_n{\bf r}} - {I}_{ik}({\bf 
H}){J}_{k}^{(n)}\right\}. 
\eqno (10.a)
$$
where $q_{ik}^{(n)}$ is the tensor $q_{ik}$ for ${\bf k} = {\bf k}_{n}$, 
and
$$
{I}_{ik}({\bf H}) = \int {\rm d}^{3}k q_{ik}({\bf H})
\delta ({\bf k}),  \eqno (10.b)
$$
$\delta ({\bf k})$ is the $\delta$-function.

We use Eq.(10.a) to present ${\delta E}_{k_{n-1}}^{(n-1)}({\bf r})$ as a 
functional of ${\delta E}_{j}^{(n-2)}$. After Fourier analyzing the 
obtained expression, we substitute it in Eq.(10.a). If in the subsequent 
order we repeat this decreasing procedure, we obtain 
$$
\sum_{l=2}^{m} v_{ik}^{(m-l)}J_{k}^{(l)} = -w_{ik}^{(m)}<E_k>, 
\eqno (11)
$$
where
$$
v_{ik}^{(0)} = {\delta}_{ik}, \quad v_{ik}^{(1)} = 0, \quad
{v}_{ik}^{(n)} = {w}_{im}^{(n)}{I}_{mk}, \; {\rm if}\; n \ge 2.
\eqno (12)
$$
With the replacement of ${\Delta}_{j_rl_{r+1}}({\bf H},{\bf k}_r- {\bf 
k}_{r+1})$ by ${\Delta}_{j_rl_{r+1}}({\bf H},{\bf r}_r)$ in the 
integrand of the expression for ${w}_{ik}^{(n)}$, we obtain
$$ 
w_{ik}^{(n)}({\bf H}) = \frac{{(-1)}^{n}}{{(2\pi)}^{3(n-1)}} 
\int...\int 
{\rm d}^{3}k_{1}...{\rm d}^{3}k_{n-1}{\rm e}^{i{\bf k}_1{\bf r}}
q_{l_1j_1}^{(1)}q_{l_2j_2}^{(2)}... q_{l_{n-1}j_{n-1}}^{(n-
1)}\times
$$
$$
\int...\int {\rm d}^{3}r_{1}...{\rm d}^{3}r_{n-1}
{\rm e}^{-i{\bf r}_1({\bf k}_1 -{\bf k}_2)}
{\rm e}^{-i{\bf r}_2({\bf k}_2 -{\bf k}_3)}...
{\rm e}^{-i{\bf r}_{n-1}{\bf k}_{n-1}}\times
$$
$$ 
<{\Delta}_{il_1}({\bf H},{\bf r}){\Delta}_{j_1l_2}({\bf H},{\bf r}_1)... 
{\Delta}_{j_{n-1}k}({\bf H},{\bf r}_{n-1})>. \eqno (13)
$$

If we introduce 
$$
W_{ik} = \sum_{m=2}^{\infty} w_{ik}^{(m)};  \quad
V_{ik} = \sum_{m=0}^{\infty} v_{ik}^{(m)} = {\delta}_{ik} + 
W_{im}I_{mk},
\eqno (14)
$$
summing the equations (11) for all $2 \le m < \infty$, we have
$$
J_i = -{V}_{ij}^{-1}W_{jk}<E_k>.  \eqno (15)
$$
Consequently, according to Eq.(5) 
$$
{\sigma}_{ik}^{ef}= <{\sigma}_{ik}({\bf H})> - 
{V}_{ij}^{-1}({\bf H})W_{jk}({\bf H}). \eqno (16)
$$

If for a given uniform magnetic field we know LCT and the inhomogeneity 
of the metal is not very strong (the perturbation series converge), 
Eq.(16) allows us to calculate ECT of an inhomogeneous metal with an 
arbitrary accuracy with respect to inhomogeneity. However, to calculate 
the many-point correlators in the integrand of Eq.(13), we need to 
define certain statistical properties of our inhomogeneous medium. 
Usually, the dependence on the statistical properties manifests itself, 
when after calculation of the contractions of mani-point averages with 
the tensors $q_{ik}$ (see Eq.(7.b)) in the right-hand side of Eq.(13), 
scalar products ${\bf k}_{p}{\bf k}_{q}$ ($p \ne q$) enter the 
integrand.

In the next Section we show that sometimes even in the 3-D case summing 
the series (14), we can calculate some elements of ECT exactly. For 
example, this is true for the longitudinal conductivity of
polycrystals of metals with closed Fermi surfaces in high magnetic 
fields.

\section{Exact solution for the effective conductivity of polycrystals 
in high magnetic fields}

Polycrystals are widespread case of inhomogeneous media, where the 
inhomogeneity is due to different orientation of discrete single crystal 
grains. If crystallographic axes of the grains are randomly rotated with 
respect to a fixed set of laboratory axes, the elements of LCT measured 
in the laboratory coordinate system are stochastic functions of 
position. When calculating the conductivity tensor in a uniform magnetic 
field, it is natural to connect the laboratory axes with the direction 
of the vector $\bf H$. In what follows we assume that the magnetic 
vector is directed along the axis 3 of the laboratory coordinate system.

We examine polycrystals of metals with closed Fermi surfaces supposing 
that number of conduction electroons $n_e$ is not equal to the number of 
holes $n_h$. We examine the case of very high magnetic fields ($r_c \ll 
l$). Under these suppositions the invariant form of the asymptotic 
expression for LCT ${\sigma}_{ik}({\bf H},{\bf r})$ can be written as 
(see, e.g., \cite{12})
$$
{\sigma}_{ik}({\bf H},{\bf r}){|}_{H \to \infty} = 
S(\vec\kappa , \vec r ){\kappa}_{i}{\kappa}_{k} + 
\frac{1}{H}{e}_{ikl}{a}_{lm}(\vec\kappa , \vec r){\kappa}_{m} + 
\frac{1}{H^2}{s}_{ik}(\vec\kappa , \vec r), \eqno (17)
$$
where ${\kappa}_{i} = H_i/H$ (in the laboratory coordinate system 
$\vec\kappa = (0,0,1)$) and ${s}_{ik} = {s}_{ki}$. We also suppose that 
due to the symmetry of our single crystal ${a}_{ik} = {a}_{ki}$. The 
elements of the tensors ${s}_{ik}$, ${a}_{ik}$ as well as the scalar $S$ 
depend on the fixed vector $\vec\kappa = \vec H/H$. The expression (17) 
is the expansion of of the elements of the tensor ${\sigma}_{ik}({\bf 
H},{\bf r})$ in powers of the small parameter $r_c/l$: the first term 
corresponds to the leading term of the longitudinal local conductivity 
that is independent of the parameter $1/H$, the second term is the Hall 
conductivity, and the term proportional to $1/H^2$ contributes to 
the transverse conductivity. 

Our goal is to calculate the leading terms of the elements of ECT. This 
means that we calculate the longitudinal effective conductivity up to 
the terms independent of the parameter $1/H$. We will show that 
fluctuating part of LCT does not enter the effective Hall conductivity. 
Also our calculations show that the transverse effective conductivity 
cannot be calculated with the aid of the approach used when calculating 
the longitudinal and the Hall effective conductivites. The difficulties 
one faced when calculating the transverse effective conductivity are 
clearly seen from the results of \cite{15}.

For our calculations it is important that in the laboratory coordinate 
system the local element ${\sigma}_{12}^{(H)}$ of the Hall conductivity 
is defined by the difference $n_e - n_h$ only: ${\sigma}_{12}^{(H)} = 
1/{R}_{\infty}H $, where ${R}_{\infty} = 1/ec(n_e - n_h)$ is the Hall 
constant. In the invariant notation ${\sigma}_{12}^{(H)} = 
{a}_{ik}{\kappa}_{i}{\kappa}_{k}$. Thus, for all the grains the 
contraction ${a}_{ik}({\bf H},{\bf r}){\kappa}_{i}{\kappa}_{k} = 
<{a}_{ik}{\kappa}_{i}{\kappa}_{k}> = 1/{R}_{\infty}$.

The ensemble averages $<{s}_{ik}>$ and $<{a}_{ik}>$ are symmetric 
isotropic second rank tensors whose elements depend on $\vec\kappa$. The 
general form of a symmetric isotropic tensor $M_{ik}(\vec\kappa)$ is 
$$
M_{ik}(\vec\kappa)= M_1{\delta}_{ik} + M_2{\kappa}_{i}{\kappa}_{k},
$$
$$
M_1 = \frac{1}{2}[{M}_{kk}-{M}_{ik}{\kappa}_{i}{\kappa}_{k}], 
\quad
M_2 = \frac{1}{2}[3{M}_{ik}{\kappa}_{i}{\kappa}_{k}-{M}_{kk}]. \eqno 
(18)
$$
Consequently, if we set $<{s}_{ik}> = S_1{\delta}_{ik} + 
S_2{\kappa}_{i}{\kappa}_{k}$ and $<{a}_{ik}> = A_1{\delta}_{ik} + 
A_2{\kappa}_{i}{\kappa}_{k}$, we have
$$
S_1 = \frac{1}{2}[<{s}_{kk}>-<{s}_{ik}>{\kappa}_{i}{\kappa}_{k}], 
\quad
S_2 = \frac{1}{2}[3<{s}_{ik}>{\kappa}_{i}{\kappa}_{k}-
<{s}_{kk}>].\eqno (19.a)
$$ 
and
$$
A_1 = \frac{1}{2}[<{a}_{kk}>-1/{R}_{\infty}], \quad
A_2 = \frac{1}{2}[3/{R}_{\infty}-<{a}_{kk}>].\eqno (19.b)
$$
When writing Eq.(19.b) we used the equality  
$<{a}_{ik}{\kappa}_{i}{\kappa}_{k}> = 1/{R}_{\infty}$.

Thus, averaging LCT (17) we obtain
$$
<{\sigma}_{ik}({\bf H})>{|}_{H \to \infty} = <S(\vec\kappa)> 
{\kappa}_{i}{\kappa}_{k} + 
\frac{1}{H{R}_{\infty}}{e}_{ikm}{\kappa}_{m}
+ \frac{1}{H^2}[S_1{\delta}_{ik} + S_2{\kappa}_{i}{\kappa}_{k}].
\eqno (20)
$$
If ${s}_{ik} = <{s}_{ik}> + \delta {s}_{ik}$, ${a}_{ik} = <{a}_{ik}> + 
\delta 
{a}_{ik}$ and $S = <S> + \delta S$ ($<\delta {s}_{ik}> = <\delta 
{s}_{ik}> = 
<\delta S > =0$), the fluctuating part of the LCT is
$$
{\Delta }_{ik}({\bf H},{\bf r}){|}_{H \to \infty} = \delta 
S(\vec\kappa , \vec r){\kappa}_{i}{\kappa}_{k} + 
\frac{1}{H}{e}_{ikl}{\delta a}_{lm}(\vec\kappa , \vec r){\kappa}_{m} + 
\frac{1}{H^2}{\delta s}_{ik}(\vec\kappa, \vec r). \eqno (21.a)
$$
With regard to the abovementioned argumentation, the contraction
$$
{\delta a}_{ik}{\kappa}_{i}{\kappa}_{k} = 0. \eqno (21.b)
$$

To make use of Eq.(16), we first calculate the terms $w_{ik}^{(n)}$ of 
the series $W_{ik}$ (see Eq.(15)). Let us write down the product of the 
tensors 
${\Delta}_{il_1}({\bf H},{\bf r}){\Delta}_{j_1l_2}({\bf H},{\bf r}_1)... 
{\Delta}_{j_{n-1}k}({\bf H},{\bf r}_{n-1})$ entering the 
integrand of the right-hand side of Eq.(13). Having in mind to calculate 
the effective longitudinal and the Hall conductivities only, we write 
this product up to the terms of the order of $1/H$. Then
$$
{\Delta}_{il_1}({\bf H},{\bf r}){\Delta}_{j_1l_2}({\bf H},{\bf r}_1)... 
{\Delta}_{j_{n-1}k}({\bf H},{\bf r}_{n-1}) = ({\hat L}^{(n)} + 
\frac{1}{H}{\hat A}^{(n)} + ...){\hat P}^{(n)}, 
\eqno (22)
$$
where the hat above the letter denotes a set of subscripts. In this 
notations ${\hat P}^{(n)}$ is the product
$$
{\hat P}^{(n)}  = \prod_{q=1}^{n-1}{\kappa}_{l_q}{\kappa}_{j_q}. 
\eqno (23)
$$

In Eq.(22) the tensor ${\hat L}^{(n)}$ appears from the product of $n$ 
tensors 
$\delta S({\bf r}_q){\kappa}_{j_q}{\kappa}_{l_{q+1}}$: 
$$
{\hat L}^{(n)} = {\kappa}_{i}{\kappa}_{k}
\delta S({\bf r})\prod_{p=1}^{n-1}\delta S({\bf r}_p).
\eqno (24)
$$
In what follows we omit the fixed vector $\vec\kappa$ in the arguments 
of the stochastic functions.

The term ${\hat A}^{(n)}$ corresponds to all possible products including 
one  tensor ${\delta a}_{mn}({\bf r}_s)$ and $n-1$ tensors 
$\delta S({\bf r}_q){\kappa}_{j_q}{\kappa}_{l_{q+1}}$. When $n=2$,
${\hat A}^{(2)} = {\kappa}_{i}{\kappa}_{l_1} {e}_{j_1km}\delta S({\bf 
r})
{\delta a}_{mn}({\bf r}_1){\kappa}_{n} + {\kappa}_{j_1}{\kappa}_{k} 
{e}_{il_1m}\delta S({\bf r}_1){\delta a}_{mn}({\bf r}){\kappa}_{n}$. If 
$n>2$,
$$
{\hat A}^{(n)} = {e}_{il_1m}\frac{{\kappa}_{k}}{{\kappa}_{l_1}}
{\delta a}_{mn}({\bf r}){\kappa}_{n}\prod_{p=1}^{n-1}\delta S({\bf r}_p) 
+
{e}_{j_{n-1}km}\frac{{\kappa}_{i}}{{\kappa}_{j_{n-1}}}
\delta S({\bf r}){\delta a}_{mn}({\bf r}_{n-1}){\kappa}_{n}
\prod_{p=1}^{n-2}\delta S({\bf r}_p) +
$$
$$
+ {\kappa}_{i}{\kappa}_{k}\sum_{s=1}^{n-2}{e}_{j_sl_{s+1}m}
\frac{1}{{\kappa}_{j_s}{\kappa}_{l_{s+1}}}
\delta S({\bf r}){\delta a}_{mn}({\bf r}_{s}){\kappa}_{n}
\prod_{p=1;p\ne s}^{n-1}\delta S({\bf r}_p). \eqno (25)
$$

When calculating the tensor $w_{ik}^{(n)}({\bf H})$, we also need to 
know  
the tensors $q_{l_sj_s}^{(s)}$ entering the integrand of Eq.(13). Taking 
account of the expression for the averaged conductivity, we see that 
the contraction $<{\sigma}_{ik}({\bf H})>k_i^{(s)}k_k^{(s)}/k^2_s = 
(<S> + S_2/H^2){(\vec\kappa{\vec k}_{s}/k_s)}^2 + S_1/H^2$. Then, in 
high 
magnetic fields in the leading approximation 
$$
q_{l_sj_s}^{(s)} = \frac{k_{l_s}^{(s)}k_{j_s}^{(s)}}{{(\vec\kappa
{\vec k}_{s})}^2<S>}. \eqno (26.a)
$$
Consequently, the leading term of the contraction 
$q_{l_sj_s}^{(s)}{\kappa}_{l_s}{\kappa}_{j_s} = 1/<S>$, and the 
contraction
$$
X^{(n)} = q_{l_1j_1}^{(1)}q_{l_2j_2}^{(2)}\dots q_{l_{n-1}j_{n-1}}^{(n-
1)}
{\hat P}^{(n)} = \frac{1}{{<S>}^{n-1}}. \eqno (26.b)
$$
The tensor ${\hat P}^{(n)}$ is defined by Eq.(23).

Let us calculate the contributions of different terms in Eq.(22) to the 
elements of the tensor $w_{ik}^{(n)}$. We set $w_{ik}^{(n)} = 
w_{ik}^{(n;L)} + w_{ik}^{(n;A)}/H $. The tensors $w_{ik}^{(n;L)}$, 
$w_{ik}^{(n;A)}$ are defined by the products 
${\hat L}^{(n)}{\hat P}^{(n)}$ and ${\hat A}^{(n)}{\hat P}^{(n)}$, 
respectively. 

We start with calculation of $w_{ik}^{(n;L)}$. Substituting the product 
${\hat L}^{(n)}{\hat P}^{(n)}$ in Eq.(13), we see that only the 
contraction $X^{(n)}$ enters the integrand in the right-hand side of 
this equation. Then
$$ 
w_{ik}^{(n;L)} = \frac{{(-1)}^{n}{\kappa}_{i}{\kappa}_{k}}
{{<S>}^{n-1}{(2\pi)}^{3(n-1)}} \int...\int 
{\rm d}^{3}k_{1}...{\rm d}^{3}k_{n-1}{\rm e}^{i{\bf k}_1{\bf r}}
\int...\int {\rm d}^{3}r_{1}...{\rm d}^{3}r_{n-1} \times
$$
$$
{\rm e}^{-i{\bf r}_1({\bf k}_1 -{\bf k}_2)}
{\rm e}^{-i{\bf r}_2({\bf k}_2 -{\bf k}_3)}...
{\rm e}^{-i{\bf r}_{n-1}{\bf k}_{n-1}}
<\delta S({\bf r})\prod_{p=1}^{n-1}\delta S({\bf r}_p)>. 
\eqno (27.a) 
$$
After the integration over all the vectors ${\bf k}_j$ ($j = 1, 2,...,n-
1$), the 
$\delta$-functions $\delta ({\bf r}_{j} - {\bf r}_{j-1})$ appear in the 
integrand. These $\delta$-functions allow us to perform integration over 
position vectors ${\bf r}_j$ too. As a result we obtain
$$
w_{ik}^{(n;L)} = w_{||}^{(n)}{\kappa}_{i}{\kappa}_{k}, \quad
w_{||}^{(n)} =  {(-1)}^{n}\frac{<{(\delta S)}^n>}{{<S>}^{n-1}}, 
\eqno (27.b)
$$
where $<{(\delta S)}^n>$ stands for the one-point average 
$<{(\delta S({\bf r}))}^n>$.

Before presenting the results for $w_{ik}^{(n;A)}$, 
let us briefly discuss some statistical properties of polycrystals we 
need when calculating these tensors. For details see Refs. \cite{11,16}.

The only property of the polycrystalline medium that affects the 
ensemble averages is the rotations of the crystallographic axes of the 
grains. Then ensemble average becomes the average over all possible 
rotations of the crystallites. If in the ensemble the rotations of 
different grains are statistically independent, when calculating the 
two-point average $<a({\bf r})b({\bf r}_1)>$ of random physical 
quantities $a({\bf r})$ and $b({\bf r})$, there are two cases to 
consider: 1)$\bf r$ and ${\bf r}_{1}$ are in the same grain, and 
2) $\bf r$ and ${\bf r}_{1}$ are in different grains. We denote by 
$W_2([{\bf r},{\bf r}_{1}])$ the probability of the case 1. Then $1-
W_2$ is the probability of the case 2. Evidently, $W_2([{\bf r},{\bf 
r}_{1}]) = 
1$, when ${\bf r}={\bf r}_{1}$.  Since in the case 2 the two-point 
average 
$< a({\bf r})b({\bf r}_1)> = < a({\bf r})><b({\bf r})> $, we have
$$
< a({\bf r})b({\bf r}_1)>  = < ab>W_2 + (1-W_2)<a><b>, \quad 
<ab> = <a({\bf r})b({\bf r})>. \eqno (28.a)
$$
If $a({\bf r})$ and $b({\bf r})$ are zero-mean quantities ($<a({\bf r})> 
= <b({\bf r})> =0$), the second term in Eq.(29.a) vanish.

When calculating the three-point average $< a({\bf r})b({\bf r}_1)
c({\bf r}_2)>$, let us denote by $W_3([{\bf r}_a,{\bf r}_b],{\bf r}_c)$ 
the joint conditional probability for the vectors ${\bf r}_a$ and 
${\bf r}_b$ to get in the same grain, and, simultaneously, for the 
vector ${\bf r}_c$ to get in some other grain. The probability 
$W_3([{\bf r}_a,{\bf r}_b],{\bf r}_c)$ excludes the possibility for all 
the three vectors to be in the same grain. Next, by $W_3([{\bf 
r}_{a},{\bf r}_{b},{\bf r}_{c}])$ we denote the probability for all the 
three vectors to get in the same grain. Then 
$$
< a({\bf r})b({\bf r}_1)c({\bf r}_2)> = <a><bc> W_3([{\bf r}_1,{\bf 
r}_2],
{\bf r}) + <b><ac> W_3([{\bf r},{\bf r}_2],{\bf r}_1) +
$$
$$
<c><ab> W_3([{\bf r},{\bf r}_1],{\bf r}_2) 
+ <abc> W_3([{\bf r},{\bf r}_{1},{\bf r}_{2}]); \quad <abc> = 
<a({\bf r})b({\bf r})c({\bf r})>. \eqno (28.b)
$$
If $<a({\bf r})>=<b({\bf r})>=<c({\bf r})>=0$, only the last term in the 
right-hand side of Eq.(28.b) survives. 

When calculating $w_{ik}^{(n;A)}$, let us first examine the tensor 
$w_{ik}^{(2;A)}$. According to Eq.(13), to calculate $w_{ik}^{(2;A)}$ we 
need to know the average $<{\hat A}^{(2)}{\hat P}^{(2)}> = 
{\kappa}_{i}{\kappa}_{j_1}{e}_{j_1km}<\delta S({\bf r}){\delta a}_{mn}
({\bf r}_1){\kappa}_{n}> + {\kappa}_{l_1}{\kappa}_{k} {e}_{il_1m}
<\delta S({\bf r}_1){\delta a}_{mn}({\bf r}){\kappa}_{n}>$. Since 
$<\delta S({\bf r})> = 0$ and $<{\delta a}_{mn}({\bf r})> =0$, from 
Eq.(29.a) it follows that the two-point average $<\delta S({\bf 
r}){\delta a}_{mn}({\bf r}_1){\kappa}_{n}> = <\delta S{\delta 
a}_{mn}{\kappa}_{n}>W_2([{\bf r},{\bf r}_{1}])$. Since $<\delta S{\delta 
a}_{mn}{\kappa}_{n}>$ is an isotropic vector, whose components depend on 
the fixed unit vector $\vec\kappa$, it is necessary that $<\delta 
S{\delta a}_{mn}{\kappa}_{n}>= <\delta S{\delta 
a}_{pq}{\kappa}_{p}{\kappa}_{q}>{\kappa}_{m}$. However, according to 
Eq.(21.b) the contraction ${\delta a}_{mn}{\kappa}_{m}{\kappa}_{n} = 0$. 
Thus, the average $<{\hat A}^{(2)}{\hat P}^{(2)}>=0$. With regard to 
Eq.(13) this means that the term $w_{ik}^{(2;A)}$ vanishes.

The same argumentation is valid when calculating the terms 
$w_{ik}^{(n;A)}$ for 
an arbitrary $n > 2$. Taking into account the way of calculation of the 
term 
$w_{ik}^{(n;L)}$, it is clear that in this case we are faced with 
calculation of the averages $<{(\delta S({\bf r}))}^{n-1}{\delta 
a}_{mn}({\bf 
r}_1){\kappa}_{n}>$ and $<\delta S({\bf r}_1)){(\delta S({\bf r}))}^{n-
2}
{\delta a}_{mn}({\bf r}){\kappa}_{n}>$. Because of the equality (21.b), 
these 
averages vanish. Thus, for all $n \ge 2$ the tensors $w_{ik}^{(n;A)}=0$. 
This 
also means, that the tensor $W_{ik}$ defined by Eq.(14) does not contain 
the 
term of the order of $1/H$.

According to Eq.(27.b) for all $n$ the values of $w_{||}^{(n)}$ do not 
depend on the statistical properties of the medium. This unusual 
situation is because of, first, the rather simple tensor form of 
the fluctuating part of the leading longitudinal element of LCT 
(according to Eq.(21.b) it is $\delta S(\vec\kappa , \vec 
r){\kappa}_{i}{\kappa}_{k}$), and, second, the asymptotic form of the 
tensor $q_{l_sj_s}^{(s)}$ in the limit $H \to \infty$ (see Eq.(26.a)). 
The first point defines the specific form of the product of $n$ 
fluctuating parts of LCT (see Eq.(22)). The second point makes it 
possible to cancel out the common factors ${(\vec\kappa{\vec k}_{s})}^2$ 
in the numerators and denominators of the integrands of Eq.(27.a). This 
is the reason why no correlators of physical quantities, describing the 
fluctuating part of LCT in different polycrystalline 
grains enter the integrands. The absence of the many-point correlators 
in the final expressions for $w_{||}^{(n)}$ allows us to sum the 
series and define $W_{||} = \sum_{n=2}^{\infty} w_{||}^{(n)}$. The 
result is 
$$
W_{||} = <S>\left\{<S>\left<\frac{1}{S}\right> - 1\right\}. \eqno (29)
$$
With respect to the laboratory coordinate system 
$W_{||}{\delta}_{i3}{\delta}_{k3}$ defines the longitudinal part of the 
tensor $W_{ik}$. 

According to the basic formula (16), to calculate the logitudinal 
element ECT we also need to define the tensor $V_{ik}^{-1}$. With regard 
to the definition (14), to calculate the tensor $V_{ik}$ we need to know 
tensor $I_{ik}$ (see Eq.(10.b)). This isotropic second rank tensor can 
be calculated with the aid of  
Eq.(18).  We set $I_{ik} = I_0{\delta}_{ik} + 
I_1{\kappa}_{i}{\kappa}_{k}$. By direct calculation it is easy to see 
that although each of the coefficients $I_0$ and $I_1$ is of the order 
of $H$, the sum $I_0+I_1 = \approx 1/<S>$. Then 
with respect to the laboratory coordinate system omitting the terms of 
the order of $1/H^2$ we have
$$
V_{ik} = {\delta}_{ik} +  W_{||}(I_0+I_1)]{\delta}_{i3}{\delta}_{k3}.
\eqno (30)
$$

When with the aid of Eq.(36) we calculate the reciprocal tensor 
$V_{ik}^{-1}$, our result for ECT written with respect to the laboratory 
axes up to the terms of the order of $1/H$, is
$$
{\sigma}_{ik}^{ef}({\bf H}){|}_{(r_c/l) \ll 1} = 
{\sigma}_{||}^{ef}{\delta}_{i3}{\delta}_{k3}  + 
\frac{1}{H{R}_{\infty}}{e}_{ik3}, \quad {\sigma}_{||}^{ef} = 
\frac{1}{<1/S>}. \eqno (31)
$$

The result for ${\sigma}_{||}^{ef}$ is rather obvious. Indeed, if we 
calculate ECT up to the terms independent of the small parameter $1/H$ 
only, with respect to laboratory axes LCT is ${\sigma}_{ik}({\bf H},{\bf 
r}) = S(\vec\kappa , \vec r ){\delta}_{i3}{\delta}_{k3}$. We can 
describe such a medium as a set of plane-parallel layers perpendicular 
to the direction of the magnetic field $\bf H$. It is known (see, e.g., 
\cite{1}) that in such a medium the value of the effective conductivity is 
equal to the inverse the averaged resistivity. This is just our answer for 
${\sigma}_{||}^{ef}$.

From Eq.(31) it follows that the value of the effective Hall 
conductivity is the same as in the single crystal. Thus, the fluctuation 
corrections do not manifest themselves in the effective Hall 
conductivity. From our point of view this is reasonable, since for all 
the grains the Hall conductivity written with respect to the laboratory 
coordinate system is the same, and it is equal to the Hall conductivity 
of single crystal metal.

As an example, let us write down the effective longitudinal conductivity 
for polycrystals of metals with the Fermi surface that is a uniaxial 
ellipsoid. Such a surface is the simplest example of a closed 
nonspherical Fermi surface. With respect to the crystallographic axes, 
the equation of the Fermi surface is
$$
{\varepsilon}_{F} = \frac{1}{2{m}_{\perp}}p_{\perp}^2 + 
\frac{1}{2{m}_{z}}p_{z}^2. \eqno (32)
$$
As a parameter of the anisotropy of this Fermi surface we use
$$
\nu = \frac{{m}_{\perp}}{m_z} - 1. \eqno (33)
$$
Evidently, $-1 < \nu < \infty$. We suppose that electrons are the charge 
carriers, so that $n_h = 0$ and $n_e = n$. 

In the framework of the relaxation time approximation the general 
expression for the conductivity of single crystal metals with an 
ellipsoidal Fermi surface in an arbitrary magnetic field was 
obtained in \cite{17}. From the results of \cite{17} it follows that in 
high magnetic fields the longitudinal local conductivity is 
$$
S(\vec\kappa) = \frac{{\sigma}_{z}}{1+\nu (1- \cos^2\theta)}; \quad
{\sigma}_{z} = \frac{ne^2\tau}{m_z},
$$
where $\theta$ is the angle between the principle axes $z$ of the 
ellipsoid (32) and the direction of the magnetic field $\vec\kappa$.

From Eq.(31) it follows that in this case the effective longitudinal 
conductivity is
$$
{\sigma}_{||}^{ef} = \frac{3{\sigma}_{z}(1+\nu)}{3+2\nu}. \eqno (34)
$$

The perturbation series for ${\sigma}_{||}^{ef}$ converges when 
$\delta S/<S> <1$. Direct calculations show that this inequality is 
fulfilled at least if $\nu < 1.7$.

Concluding this Section some remarks on calculation of the transverse 
part of the conductivity tensor has to be done. It can appear
that if in Eq.(22) we write the product of stochastic tensors 
${\Delta}_{il_1}({\bf H},{\bf r}){\Delta}_{j_1l_2}({\bf H},{\bf r}_1)... 
{\Delta}_{j_{n-1}k}({\bf H},{\bf r}_{n-1})$ up to the terms of the order 
of $1/H^2$, we can calculate the effective transverse conductivity too.
Indeed, the calculation similar to the one presented above, show that 
this terms define the contribution $-<{\delta a}_{mp}{\delta 
a}_{mq}{\kappa}_{p}{\kappa}_{q}>/2<S>$ to the transverse part of the 
tensor $W_{ik}$ in the basic Eq.(16). We see that this term is of the 
second order in the stochastic part of the tensor $a_{mp}$. However, as 
it follows from the results of Yu.A.Dreizin and A.M.Dykhne (see 
\cite{15}) when calculating this term we cannot restrict ourselves with 
the leading approximation for the tensor $q_{l_sj_s}^{(s)}$ only. The 
poles entering the elements of this tensor provide the contribution to 
the transverse conductivity following from the terms of the order higher 
than $1/H^2$ in the product
${\Delta}_{il_1}({\bf H},{\bf r}){\Delta}_{j_1l_2}({\bf H},{\bf r}_1)... 
{\Delta}_{j_{n-1}k}({\bf H},{\bf r}_{n-1})$. Of course, is the 
anisotropy is very small, these terms can be omitted, but this resoning 
indicates, that most likely no exact solution can be obtained for the 
transverse effective conductivity in high magnetic fields. In any case 
further analysis is necessary to obtain the correct result.

\section{Conclusions}

Concluding this paper some remarks are to the point. First,
when galvanomagnetic phenomena are investigated experimentally, the 
object of investigation is the resistivity tensor. If we know the 
leading term of the elements of ECT up to the terms of the order of 
$1/H$ exactly, we can use Eqs.(31) when calculating the 
asymptotic of the logitudinal and the Hall effective resistivities 
(${\rho}_{ik}^{ef}({\bf H}){\sigma}_{ik}^{ef}({\bf H}) = 
{\delta}_{ik}$). Then 
$$
{\rho}_{||}^{ef} = <1/S>, \quad {\rho}_{12}^{ef} = {R}_{\infty}H.
\eqno (35)
$$
We see that again the effective Hall element of ${\rho}_{ik}^{ef}({\bf 
H})$ is the same as in the single crystal. 

Second, usually, one thinks about polycrystals as of isotropic metals, 
that is as of metals with spherical Fermi surface. However, our results 
show that this idea is not justified. Indeed, for isotropic metals in 
high magnetic fields the longitudinal element of the conductivity tensor 
is ${\sigma}_{||}^{sph} = {\sigma}_0$, where ${\sigma }_0$ is the 
conductivity when $H=0$. Let us  verify this equality for polycrystal 
with ellipsoidal Fermi surface. From the results of \cite{11} it follows 
that if $\nu <1$, with high accuracy we can estimate the effective value
of ${\sigma}_0$ as $<\sigma (0)>$. Then for the case of ellipsoidal 
Fermi surface we have 
$$
\frac{{\sigma}_{||}^{ef}}{<\sigma (0)>} = \frac{9(1+\nu 
)}{(3+2\nu)(3+\nu)}.
$$
Evidently, this ratio is not equal to one.

Finally, when comparing the calculated ECT with experimental results, 
one must 
have in mind that in our calculations only the inhomogeneity due to 
different 
orientations of polycrystalline grains was taken into account. Of 
course, in 
real polycrystals there are other sources of inhomogeneity too. For 
example, we 
do not take into account the real structure of the boundaries of the 
grains. 
This simplification is justified when the grains are sufficiently large 
and the 
properties of the grain boundaries do not affect the result 
significantly.
 
\centerline{\bf ACKNOWLEDGMENTS} 

The authors are grateful to Prof. A.M.Dykhne and Prof. A.V.Chaplik for 
helpful discussions. The work of IMK was supported by RBRF Grant No. 02-
02-17226 and Scientific School Grant No. 2078.2003.2.

It is a pleasure to dedicate this paper to A.M.Dykhne, who has made so 
many significant contributions to the theory of inhomogeneous media, on 
occasion of his 70-th birthday. We wish him health, happiness, and many 
more years of fruitful work.

\end{document}